\documentclass[a4paper,11pt]{article}
\pdfoutput=1 

\usepackage{jinstpub}

\usepackage{bm}
\usepackage{algorithm2e}
\usepackage{siunitx}
\usepackage{booktabs}
\usepackage{subcaption}

\title{Principle Study of Image Reconstruction Algorithms in Muon Tomography}

\author[a,b]{Weihe Zeng,}
\author[a,b]{Xingyu Pan,}
\author[a,b,1]{Zhi Zeng,\note{Corresponding author.}}
\author[a,b]{Hao Ma,}
\author[a,b]{Ming Zeng,}
\author[a,b,c]{and Jianping Cheng}

\affiliation[a]{Department of Engineering Physics, Tsinghua University, Beijing, China}
\affiliation[b]{Key Laboratory of Particle \& Radiation Imaging (Tsinghua University), Ministry of Education, China}
\affiliation[c]{College of Nuclear Science and Technology, Beijing Normal University, Beijing, China}

\emailAdd{zengzhi@tsinghua.edu.cn}

\abstract{
  Muon tomography is a relatively new method of radiography that utilizes muons from cosmic rays and their multiple Coulomb
  scattering property to distinguish materials. Researchers around the world have been developing various detection systems and
  image reconstruction algorithms for muon tomography applications, such as nuclear reactor monitoring and cargo inspection for
  contraband. This paper studies the principle in image reconstruction of muon tomography. 
  Implementation and comparison of some popular algorithms with our simulation dataset will be presented 
  as well as some ideas of future improvements for better image qualities and material discrimination performances.
}


\begin{document}
\maketitle
\flushbottom

\section{Introduction}

Muons dominate the components of cosmic rays at the Earth's surface, the vertical intensity of which is approximately
\SI{70}{m^{-2}.s^{-1}.sr^{-1}} at sea level according to literature~\cite{Patrignani2016}. 
Researchers have studied the properties of cosmic muons including
angular and energy distribution in detail through theories and experiments~\cite{gaisser_engel_resconi_2016,Shukla2018}.
From early 1940s, theories of charged particles' interactions in medium have been established and improved continuously by later researches~\cite{Rossi1941,Moliere1948,Highland1975,Lynch1991}.
Based on these fundamental concepts, two approaches of imaging objects by measurement of cosmic muons have been invented.

The first approach measures the transmission probabilities of muons through thick objects from different directions which implies the
information about absorption in materials. This method has been successfully applied in experiments that monitor the internal structures of
large objects like volcanoes~\cite{Procureur2018}. Though muons with high kinetic energy will mostly penetrate common size objects like
vehicles and cargoes, the scattering angles between incoming and outgoing rays follow approximately Gaussian distribution with zero mean
and variance that is related to material properties~\cite{Rossi1941}. The second approach, which is named muon tomography, utilizes the scattering information and
combines theoretical predictions to obtain the internal structures of objects. This paper will study the principle in muon tomography, which
may be an promising alternative to X-ray computed tomography (CT) for security checks.

The idea of muon tomography was first developed in Los Alamos National Laboratory (LANL)~\cite{Borozdin2003,Schultz2003} and then adopted by other groups
around the world~\cite{Pesente2009,Wang2012,Jonkmans2013,Clarkson2014}. The main challenge in the design of such system is the acquisition
of muon trajectories before and after penetrating the target. Existing systems consists of multiple layers of large two dimensional position
sensitive detectors at both side of sample area. Considering the balance between position precision and economic cost, drift chambers,
multi-gap resistive plate chamber (MRPC), gas electron multiplier (GEM) and plastic scintillating fibers were selected as detectors in
different experiment facilities~\cite{Schultz2003,Pesente2009,Wang2012,Jonkmans2013,Clarkson2014}. For example, the Tsinghua University MUon
Tomography facilitY (TUMUTY) was built with 6 layers of MRPC detectors and the incoming and outgoing directions of muons are calculated by
fitting 3 interacting points of top and bottom detector groups~\cite{Wang2012}. 
Some ideas of measuring muon momentum by time-of-flight (TOF) or multiple Coulomb scattering (MCS) in dense detectors were surveyed in literature~\cite{Anghel2015,Luo2016}.

Muon tomography has the advantage of not requiring complex particle source as in X-ray or proton CT. 
However, low flux of cosmic muons limits the amount of observed data, which is a major challenge in muon tomography. 
Though discrimination of high atomic number materials such as fissile materials is proved to be applicable, 
there is still not much progress for common materials with low or medium atomic numbers. 
In addition to detection system development, image reconstruction algorithms for muon tomography would be effective approaches to investigate the opportunities. 

In this paper, we categorize algorithms for muon tomography into two stages, muon trajectory estimation and image reconstruction.
Since the exact trajectories of muons inside imaging objects are unknown, estimation algorithms should be applied to measurements,
which include the simple straight line path (SLP), point of closest approach (PoCA)~\cite{Schultz2003} and most likely path (MLP) from MCS theory~\cite{Schulte2008,Yi2016,Chatzidakis2018}.
Tomography results are usually described as three dimensional voxelized images that each voxel contains some unique material properties.
For muon tomography, the materials property is called scattering density, which is estimated from the variance of scattering angles in each voxel. The second stage of muon tomography aims to determine scattering densities of each voxel from accumulated muon trajectories.
Various algorithms have been proposed for the second stage, including assigning the total scattering angles to PoCA~\cite{Schultz2003}, statistical reconstruction based on MCS theory~\cite{Schultz2007a} and
commonly used algebraic reconstruction technique algorithm (ART) in X-ray CT~\cite{Liu2018}.
Universal algorithms that improve the radiological image qualities have also been tested such as total variation (TV) regularization~\cite{Yu2016}.

In Section \ref{sec:method}, details of muon tomography methodologies and various algorithms will be reviewed. Test results of some
algorithms with simulation dataset will be shown in Section \ref{sec:results}. The final section will summarize these algorithms and
discuss inspirations from our test results. Future improvements of algorithms will make muon tomography more practical and compatible in industrial applications.

\section{Theory and Algorithms}\label{sec:method}

\subsection{Multiple Coulomb Scattering}

According to the scattering theory of charged particles from Rossi~\cite{Rossi1941}, charged particles traversing a plate
of medium by MCS have scattering angles and lateral displacements that follow Gaussian distribution.
Derived from the marginal distribution of scattering angles, the variance is the well known Rossi formula.

\begin{equation}
  \label{eqn:rossi}
  \sigma_\theta^2 = \dfrac{E_s^2 }{2 p^2 v^2}\cdot \dfrac{x}{X_0}
\end{equation}

In the formula, $\sigma_\theta^2$ is the variance of scattering angles, $E_s$ is a constant with a value of \SI{21}{MeV}, $X_0$ is the radiation length of materials, $x$ is the thickness of plate,
$p$ and $v$ are the momentum and velocity of muons respectively.

Moliere improved the theory of MSC and introduced more complex formula for scattering angles. Highland observed that experimentalists
preferred the simple formula from Rossi~\cite{Highland1975}. Thus, Highland added a correction term in the formula by fitting results
from Moliere's theory and suggested a new value of $E_s$ constant to be \SI{17.5}{MeV}.

\begin{equation}
  \sigma_\theta^2 = \dfrac{E_s^2 }{2p^2 v^2}\cdot \dfrac{x}{X_0} (1 + 0.125 \log_{10}\dfrac{x}{0.1X_0} )^2
\end{equation}

Lynch and Dahl further improved the approximation of scattering angle distribution by alternating some constants in Highland's formula
~\cite{Lynch1991}. We have the relatively simple and accurate formula for variance of MCS presented in many research articles.

\begin{equation}
  \label{eqn:lynch}
  \sigma_\theta^2 = \left(\dfrac{13.6~\mathrm{MeV}}{p v}\right)^2 \cdot \dfrac{x}{X_0} (1 + 0.088 \log_{10}\dfrac{x}{X_0} )^2
\end{equation}

Scattering density was derived from Rossi formula (\ref{eqn:rossi}) for material discrimination in muon tomography.
With known muon penetration thickness, we define scattering density ($\lambda$) to be a value that depends on radiation length and muon momentum.

\begin{equation}
  \lambda \triangleq \dfrac{\sigma_\theta^2}{x} = \dfrac{15~\mathrm{MeV}}{p^2 v^2}\cdot \dfrac{1}{X_0}
\end{equation}

\begin{figure}[!htb]
  \centering
  \includegraphics
    [width=0.8\hsize]
    {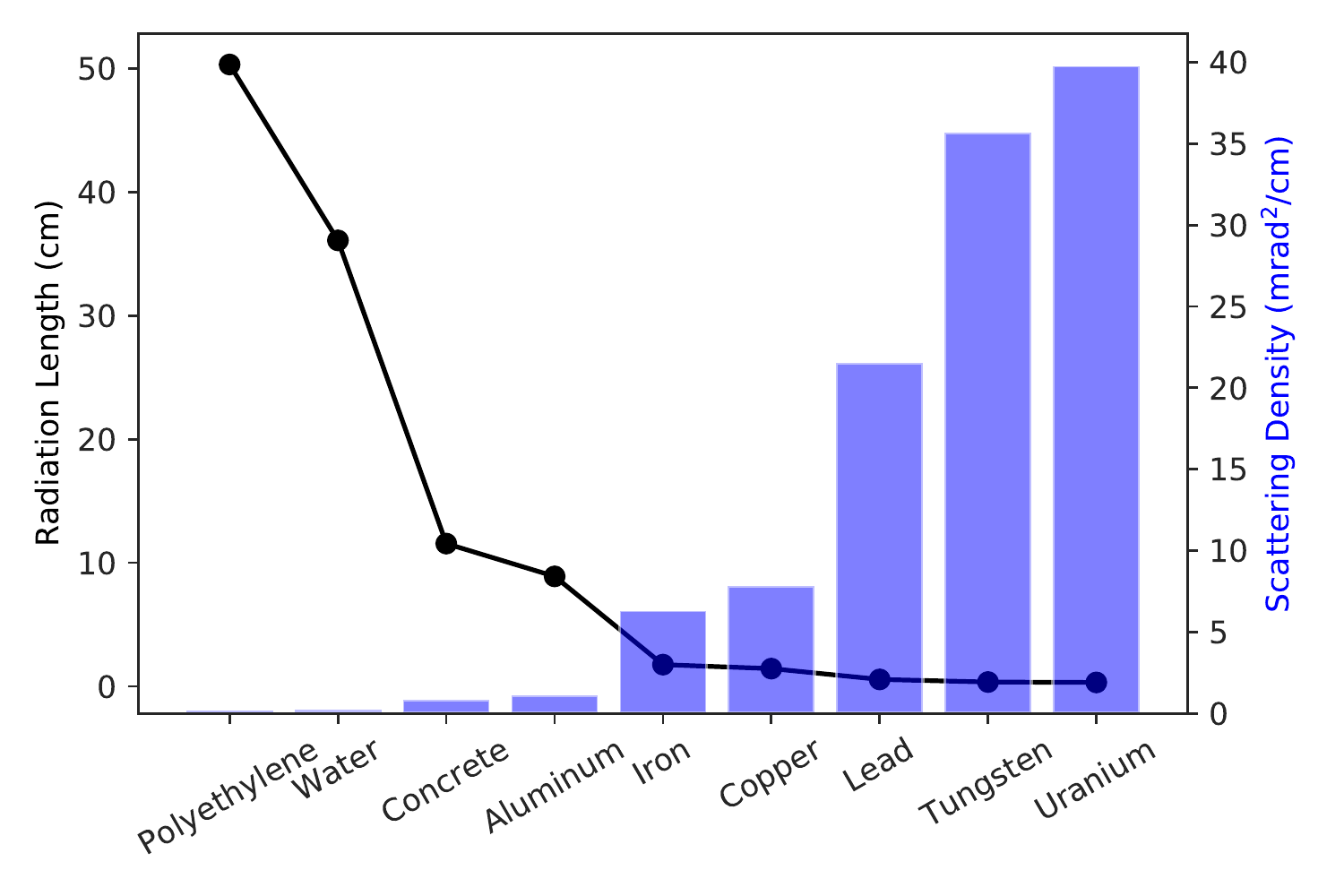}
  \caption{Radiation lengths and scattering densities of materials at 4 GeV}
  \label{fig:scatdensity}
\end{figure}

Therefore, the procedure of muon tomography can be simply described as accumulating scattering angles at different locations of the
imaging object and then generating a map of scattering densities based on their variance calculation. Figure~\ref{fig:scatdensity} shows
the radiation lengths and scattering densities of some materials and demonstrates the capability of discrimination by muon tomography.

\subsection{Trajectory Estimation}

In muon tomography, detection system is designed to measure incoming and outgoing muon trajectories as accurate as possible.
However, trajectories of muons passing through the interior of imaging objects remain a ``black box''.
Although actual trajectories are quite complicated and unpredictable, some estimation algorithms exist based on the small scattering angle nature of MCS.

\subsubsection{Straight Line Path}

Usually, the imaging objects are thin compared to the high penetration ability of cosmic muons,
and the MCS theory suggests that the distribution of scattering angles centers at 0.
One obvious method of obtaining the trajectory is to connect the entry point and exit point directly with a straight line.
In voxelized space, SLP can be efficiently calculated using Siddon algorithm~\cite{Siddon1985,Jacobs1998}.

This algorithm is accurate enough and widely used for X-ray CT,
while divergence of trajectories is not negligible in charged particle related tomography.
Other estimation algorithms taking scattering into account show superior imaging quality when applied in muon tomography.

\subsubsection{Point of Closest Approach}

As the name suggests, PoCA is the point on the closest approach from one line to the other.
In the 2 dimensional space, PoCA is simply the intersection between two unparalleled lines.
While unparalleled lines in 3 dimensional space do not always intersect, closest approach is the line
segment that is simultaneously perpendicular to both lines.
In practice, PoCA is defined to be the center of closest approach which can be obtained by solving the following equations.

\begin{equation}
  \label{eqn:poca}
  \begin{cases}
    & P_1 = \bm{p}_1 + t_1 \bm{d}_1 \\
    & P_2 = \bm{p}_2 + t_2 \bm{d}_2 \\
    & \overrightarrow{P_1 P_2} \cdot \bm{d}_1 = 0 \\
    & \overrightarrow{P_1 P_2} \cdot \bm{d}_2 = 0 \\
  \end{cases}
\end{equation}
where $P = \bm{p} + t \bm{d}$ is the parametric equation of points on straight lines,
with $\bm{p}_1, \bm{d}_1$ and $\bm{p}_2, \bm{d}_2$ being the known points and directions of incoming
and outgoing muon trajectories respectively.

After the derivation of PoCA, the estimated trajectories is a polyline composed of entry point, PoCA and exit point. This algorithm is a better approximation that SLP, but a portion of muon trajectories will be discarded when their PoCA is not in the object space.

\subsubsection{Most Likely Path}

Original developed in proton tomography researches, MLP combines MCS theory and maximum likelihood estimation to find the particle trajectories.
The likelihood of muon passing point given the information of entry and exit points is formulated by Schulte~\cite{Schulte2008}.

\begin{equation}
  \label{eqn:mpt}
  L(Y | Y_1) = L(Y_1|Y) L(Y |Y_0)
\end{equation}

$Y$ in the equation represents a vector containing scattering angle and lateral displacement at depth in the medium, while $Y_0$ and $Y_1$ are directions and displacements of entry and exit points
From MCS theory, the scattering angle and lateral displacement follow a joint Gaussian distribution. 
The covariance matrix of such distribution is shown in Eq.~\ref{eqn:mptsigma} with $\lambda$ and $z$ being the scattering density and depth in the medium.

\begin{equation}
  \label{eqn:mptsigma}
  \Sigma = \lambda \begin{bmatrix}
                      \dfrac{z^3}{3} & \dfrac{z^2}{2} \\
                      \dfrac{z^2}{2} & z
                    \end{bmatrix}
\end{equation}

\begin{figure}[!htb]
  \centering
  \includegraphics
    [width=0.6\hsize]
    {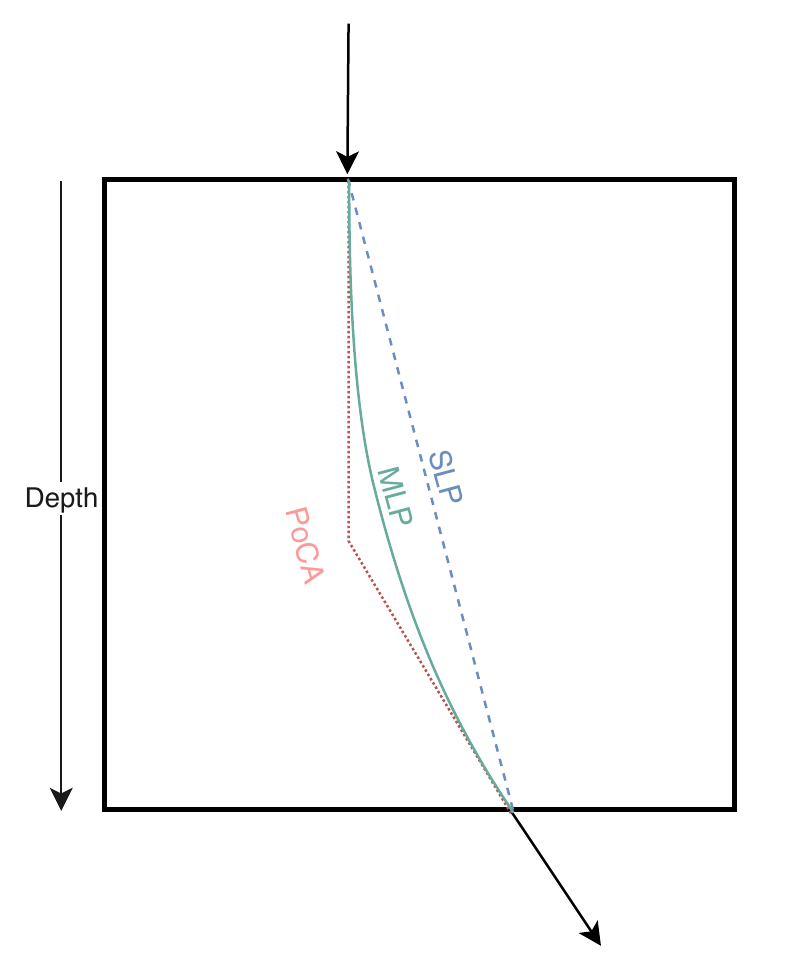}
  \caption{Illustration of trajectory estimation algorithms}
  \label{fig:mlp}
\end{figure}

The expression of MLP is derived by maximizing the probability of muon passing point $Y$ as shown in Eq.~\ref{eqn:mlp}, where $R_0$ and $R_1$ are matrix that translate entry and exit points to
corresponding depth, and covariance matrix $\Sigma_0$ and $\Sigma_1$ are also calculated at that depth.

\begin{equation}
  \label{eqn:mlp}
  Y_\text{MLP} = (\Sigma_0^{-1} + R_1^T \Sigma_1^{-1} R_1)^{-1} (\Sigma_0^{_-1}R_0Y_0 + R_1^T\Sigma_1^{-1} Y_1)
\end{equation}

MLP represents the statistical most probable trajectory of muons that entering and exiting at
the same points and angles. For charged particles that energy loss in the medium is not negligible,
the elements of covariance matrix become integral expressions to include this factor~\cite{Chatzidakis2018}. Since the prior of scattering densities in the object is not known,
homogeneous medium is often assumed that will smooth the actual trajectory of muon.

\subsection{Reconstruction}

With trajectory estimation algorithms and accumulation of cosmic muon measurements, the next stage requires an algorithm that 
generates the distribution of scattering densities inside the imaging object. 
Researchers have developed various image reconstruction algorithms for muon tomography and compared them with the work of predecessors
in terms of resolution, noise-to-signal ratio, etc. 

\subsubsection{Direct Allocation}

By definition, scattering density is the variance of scattering angles scaled by traversing thickness. 
Thus, images can be reconstructed after each voxel has enough data accumulation of scattering angles. 
Direct allocation methods analyze the estimated muon trajectories and assign scattering angles directly to some voxels. 

The simplest strategy would be assigning the average scattering angle to each voxel along the path. 
This strategy may be useful for homogeneous medium, however, structures with large scattering density differences are our target of interest. 
The most commonly used direct allocation method utilizes the PoCA information and assumes that the single point accounts for the total scattering 
while other points in the trajectory have zero scattering~\cite{Schultz2003}. 
This method is often called PoCA algorithm in literatures and should be distinguished from PoCA in trajectory estimation. The reconstruction process is shown in Algorithm~\ref{alg:poca}. 

\begin{algorithm}[!htb]
  \KwData{Muon trajectories}
  \KwResult{Scattering densities of each voxel}
  Initialize scattering angle container $S_i$ for each voxel\;
  \ForEach{muon trajectory $\mu$}{
   Calculate angle $\theta$ between incoming and outgoing muon\;
   \ForEach{voxel $v_i$ in $\mu$}{
    \eIf{PoCA in $v_i$}{
      Append $\theta$ to $S_i$ \;
      }{
      Append $0$ to $S_i$ \;
    }
   }
  }
  Calculate variance of $S_i$ for each voxel, and $\lambda_i = \dfrac{\text{Var}[S_i]}{x_i}$
  \caption{Process of PoCA algorithm}
  \label{alg:poca}
\end{algorithm}

In addition to the basic PoCA algorithm, researchers have explored some ideas of improvement. 
For example, ``pitchfork'' method was introduced considering the measurement uncertainties, 
where the muon trajectories are randomly sampled according to presumed uncertainties and scattering densities 
are evaluated on this expended dataset~\cite{Jonkmans2013}. 
Algorithm that utilizes MLP estimation was also examined by assigning scattering angles to the most distant point along MLP directly~\cite{Yi2016}. 

\subsubsection{Maximum Likelihood Scattering and Displacement}

Direct allocation methods are easy to implement and efficient in computation, while reconstructed images tend to be noisy due to 
the oversimplified assumption of scattering and uncertainties in trajectory estimation. 
Schultz developed an algorithm based on statistical model of MCS in his dissertation~\cite{Schultz2003} and named it 
Maximum Likelihood Scattering and Displacement (MLSD). 
Further improvement of maximization method in MLSD using expectation–maximization (EM) algorithm was presented in article~\cite{Schultz2007a}. 

The likelihood of scattering densities $\lambda$ in imaging object with $M$ measurement data $D$ can be expressed as Eq.~\ref{eqn:mlsd0}, in which the data is composed of both the scattering angles and lateral displacements
Each term in the production is derived from Gaussian approximation of MCS as shown in Eq.~\ref{eqn:mlsd1} with the same 
expression for covariance matrix in MLP algorithm (\ref{eqn:mptsigma}). 

\begin{equation}
  \label{eqn:mlsd0}
  P(D|\lambda) = \prod_{i\leq M} P(D_i|\lambda)
\end{equation}

\begin{equation}
  \label{eqn:mlsd1}
  P(D_i|\lambda) = \dfrac{1}{2\pi |\Sigma_i|^{1/2}} \exp\left( -\dfrac{1}{2} D_i^T \Sigma_i^{-1} D_i \right)
\end{equation}

EM iteration was proposed for solving the maximum likelihood problem in MLSD with the auxiliary function shown in (\ref{eqn:mlsdobject}),
where $H$ represents the hidden data inside imaging object.  

\begin{align}
  \label{eqn:mlsdobject}
  \lambda^{(n+1)} =~& \arg \max_{\lambda} Q(\lambda;\lambda^{(n)}) \\
  Q(\lambda;\lambda^{(n)}) =~& {\mathbb E}_{H|D,\lambda^{(n)}} [\log P(H|\lambda)]
\end{align}

MLSD requires much more computational power to reconstruct images than those direct allocation methods, but 
this iterative approach is robust and converges to images with better quality.

\subsubsection{Maximum a Posteriori and Regularizations}

Maximum a posteriori (MAP) is another statistical model frequently used in radiological tomography. 
Similar to MLSD, the probability of scattering density given the observed data is expressed according to the Bayesian theorem. 

\begin{equation}
  \label{eqn:map0}
  P(\lambda | D) = \dfrac{P(D|\lambda) P(\lambda)}{P(D)}
\end{equation}

Then, the objective function is to maximize this probability with respect to scattering density. 

\begin{equation}
  \label{eqn:map1}
  \lambda = \arg \max _{\lambda\geq 0} P(\lambda | D)
\end{equation}

Taking the logarithm of this probability results in the following form. 
If the prior of scattering density distribution $P(\lambda)$ is flat, we have the Maximum Likelihood previously discussed. 
Otherwise, the prior term can be written as some function $U(\lambda)$ scaled by $\beta$, which penalizes the log likelihood term. 

\begin{equation}
  \label{eqn:map2}
  \begin{split}
  \lambda =~& \arg \max _{\lambda\geq 0} [L(D|\lambda) + \log P(\lambda)] \\
          =~& \arg \max _{\lambda\geq 0} [L(D|\lambda) - \beta U(\lambda)]
  \end{split}
\end{equation}

The procedure of MAP algorithm is often referred as regularization in statistical image reconstruction. 
Lots of ideas to formulate the regularizing term have been explored in X-ray CT community~\cite{Zhang2018} that  
may be borrowed and applied in muon tomography for various purposes. 
One of the most popular kind of regularization term is Markov random field model based priors, which assumes that 
material properties are only related to neighboring voxels. Thus, the regularization terms are often expressed in the following equation. 

\begin{equation}
  \label{eqn:mapMRF}
  U(\lambda) = \sum_j\sum_{m\in W_j} w_{jm} \phi(\lambda_j - \lambda_m)
\end{equation}
where $W_j$ is the set of neighboring voxels of voxel $j$ and function $\phi(\Delta)$ represents the relation that two neighboring 
voxels should follow in prior knowledge. 
Common choices of function $\phi(\Delta)$ includes quadratic function and compressed sensing based $l_p$ norm, as shown in the following equations. 

\begin{align}
  \label{eqn:mapnorm0}
  \phi(\Delta) =~& \dfrac{\Delta^2}{2} \\ 
  \label{eqn:mapnorm1}
  \phi(\Delta) =~& \left(\sum_m |\Delta_m|^p\right)^{1/p}
\end{align}

These regularization terms ensure edge preserving and noise reduction during the maximization process. 
There are some successful applications of them in muon tomography that result in better reconstructed images than not regularized MLSD~\cite{Yu2016}.

\subsubsection{Algorithms in X-ray CT}

There are other common image reconstruction algorithms in X-ray CT that have been extensively studied.  
Among them filtered backprojection (FBP) and algebraic reconstruction technique (ART) are the most popular and widely implemented in practice. 
Researchers have also explored the possibility of adapting these algorithms for better reconstruction in muon tomography~\cite{Liu2018}.
They compared the scattering density to the attenuation coefficient in X-ray CT (\ref{eqn:art0}) and generated projection matrix by grouping muons with different azimuth angle values. 

\begin{equation}
  \label{eqn:art0}
  \log\dfrac{I_0}{I} = \sum_{i} d_i \mu_i,\quad  \sigma_\theta^2 = \sum_{i} d_i \lambda_i
\end{equation}
where $I$ is the intensity of X-ray, $\mu$ is the attenuation coefficient of each voxel, $d$ is the track length. 

The equation to be solved in its matrix form is shown below. And they utilized SART algorithm to solve it and acquired the reconstruction images~\cite{Liu2018}. 

\begin{equation}
  \label{eqn:art1}
  WX=P
\end{equation}
where $W$ is a matrix containing path length of each group in each voxel, $X$ vector represents the desired scattering density map and 
$P$ is the projected variances of scattering angles.  

\subsection{Summary of Algorithms}

Algorithms for two stages in muon tomography have been developing continuously. 
Any combination of algorithms within two stages produces a complete pipe line of reconstruction. 
As shown in Table~\ref{tab:summary_alg}, MLP has not yet been extensively applied in image reconstruction, 
which may be a future approach of improvement. 
Nevertheless, the fast evolving X-ray CT should be a excellent source for brand new algorithm in muon tomography. 

\begin{table}[!htb]
  \centering
  \caption{Summary of algorithms for muon tomography}
  \label{tab:summary_alg}
  \begin{tabular}{|l|l|}
    \hline
    Trajectory Estimation & Reconstruction \\
    \hline
    SLP  & Direct allocation \\
    PoCA & MLSD \\
    MLP  & MAP \\
         & FBP/ART \\
    \hline
  \end{tabular}
\end{table}

\section{Simulation and Test}\label{sec:results}

\subsection{Monte Carlo Simulation of Muon Scattering}

\begin{figure}[!htb]
  \centering
  \includegraphics
    [width=0.8\hsize]
    {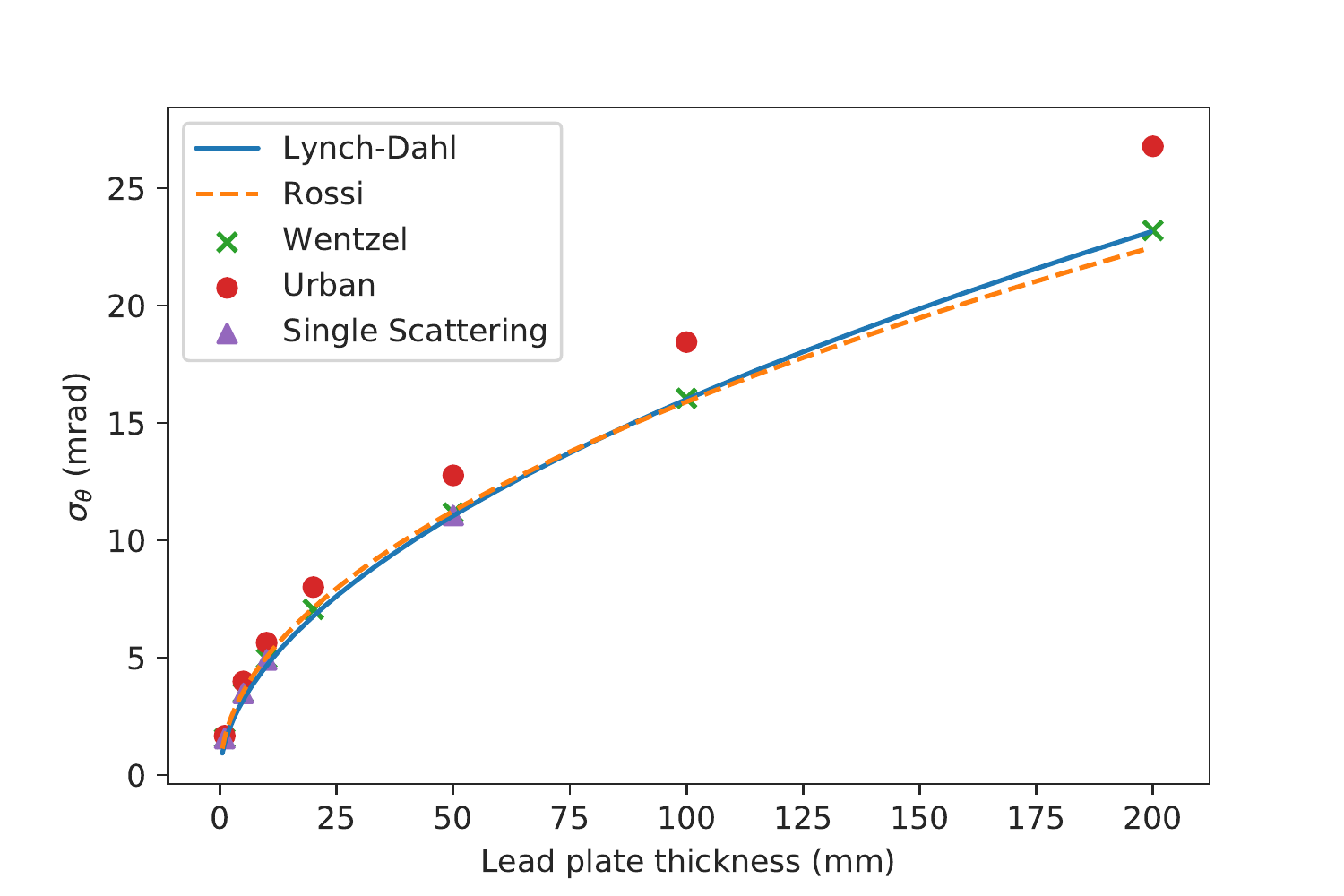}
  \caption{Comparison of GEANT4 MCS models}
  \label{fig:geant4model}
\end{figure}

Monte Carlo (MC) simulation is a powerful approach to test experiment concepts virtually based on physical models.
CERN's GEANT4 toolkit~\cite{AGOSTINELLI2003250} is one of the most popular and flexible MC platform 
that enables us to generated dataset for testing muon tomography algorithms. 
Available MCS models in GEANT4's electromagnetic processes include Wentzel Model, Urban Model and Single Scattering (SS) model. 
To validate these models with empirical formulas (\ref{eqn:rossi},\ref{eqn:lynch}) used in experiments, 
scattering angles of muons with fixed energy of 4 GeV that penetrate lead plates of different thickness were simulated with MCS models mentioned above. 
The standard deviations of simulated scattering angles were drawn alongside curves predicted by empirical formulas in Figure~\ref{fig:geant4model}. 
Single scattering should be the most accurate one since it models each occurrence of Coulomb scattering, though the computation time 
would be unacceptable when the penetrating medium is thick. 
Wentzel model agrees with Eq.~\ref{eqn:lynch} well and is chosen during the generation of simulation dataset.

\begin{figure}[!htb]
  \centering
  \includegraphics
    [width=0.8\hsize]
    {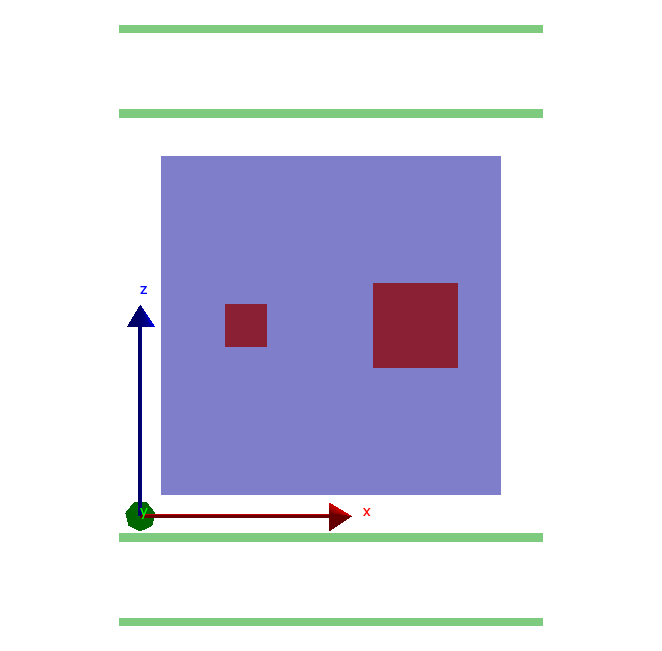}
  \caption{Geometry of MC simulation}
  \label{fig:geant4geo}
\end{figure}

Geometry setup of the simulation is shown in Figure~\ref{fig:geant4geo}. 
Two layers of position sensitive detectors (green color) is placed on each side of the sample area. 
The sample area consists of two metal cubes (red color) in a water container (blue color). 
The smaller one of the two metal cubes is made of lead, while the other one of iron. 
Other space is filled with air. 
Cosmic muons are generated by CRY~\cite{CRY2007} code, and transported through the system by GEANT4. 
Interaction points of muons and the detectors are recorded event-by-event as the simulation dataset for later image reconstruction tests. 

\subsection{Implementation of Algorithms}

\begin{figure}[!htb]
  \centering
  \includegraphics
    [width=0.9\hsize]
    {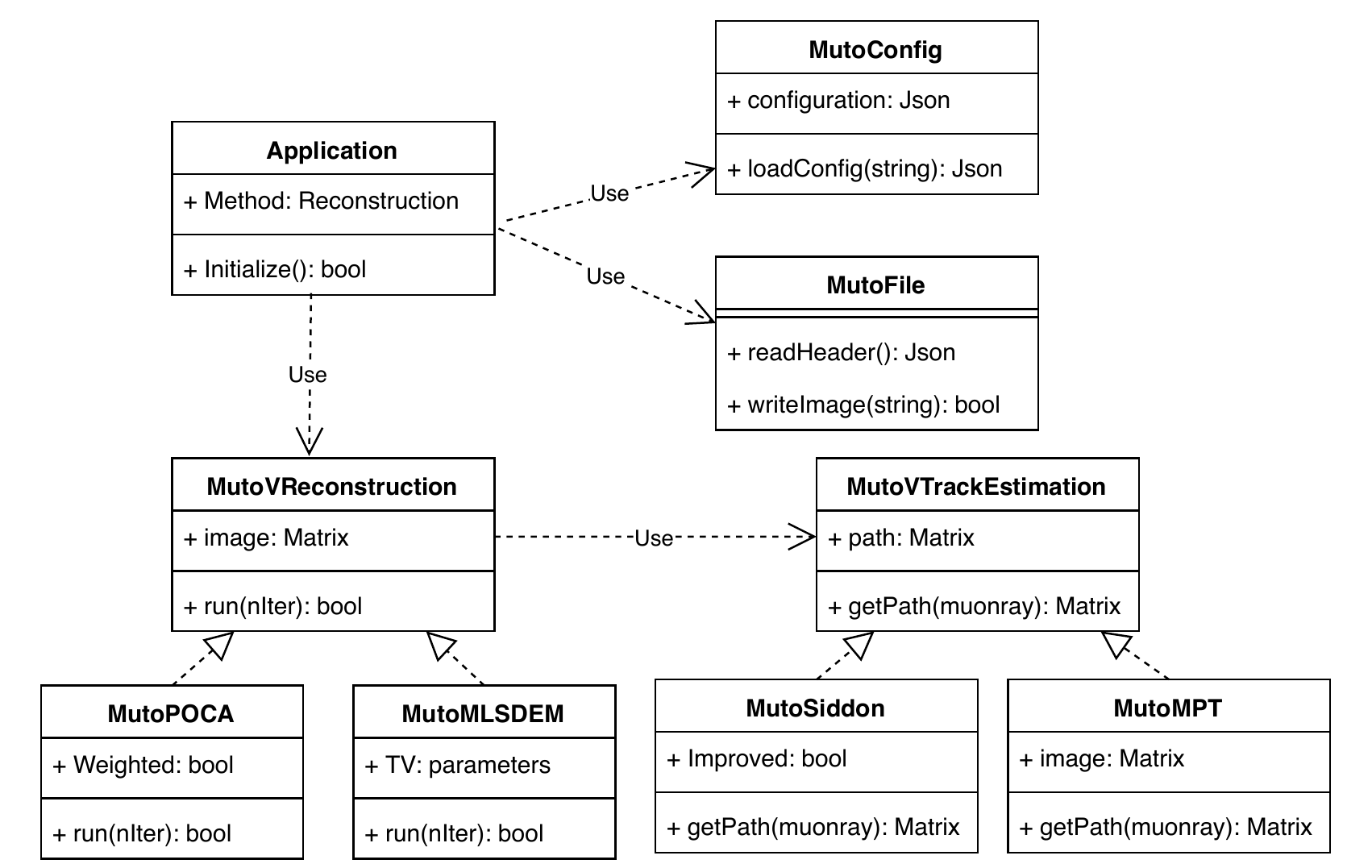}
  \caption{Design of muon tomography software}
  \label{fig:mutodesign}
\end{figure}

Review of different algorithms in previous section suggests that image reconstruction consists of two stages. 
Therefore, abstract classes for trajectory estimation and reconstruction were included by our design of software implementation 
that would enhance the flexibility and modularity of the code. 
Details of the software design is presented in Figure~\ref{fig:mutodesign}. 
Our code of muon tomography is open source and can be downloaded online\footnote[1]{https://github.com/nmtzwh/MuonTomography}. 
Currently, direct allocation with PoCA trajectory, MLSD and MAP algorithms have been implemented and tested with the simulation dataset. 
The results will be shown and discussed in the following contents. 

\subsection{Trajectory Estimation Precision}

\begin{figure}[!htb]
  \centering
  \includegraphics
    [width=0.8\hsize]
    {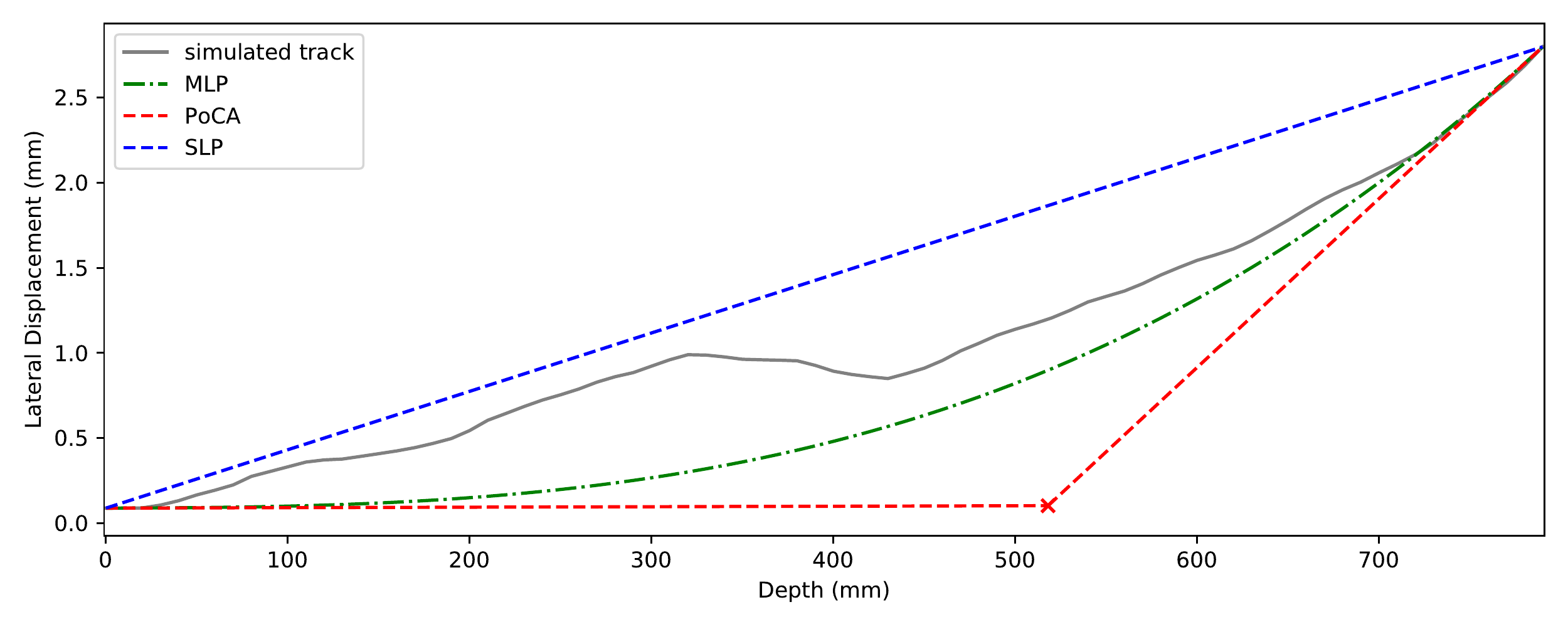}
  \caption{An example of simulated trajectory and estimated results}
  \label{fig:trackdrawing}
\end{figure}

As discussed in Section~\ref{sec:method}, there exist 3 major algorithms for charged particle trajectory estimation. 
We have simulated muons with fixed energy of 4 GeV scattering inside aluminum plate of 80 cm thickness and 
recorded their positions during the simulation as the ``true'' trajectories. 
Figure~\ref{fig:trackdrawing} shows an example of the actual path of muon against results estimated in our software. 

\begin{figure}[!htb]
  \centering
  \includegraphics
    [width=0.8\hsize]
    {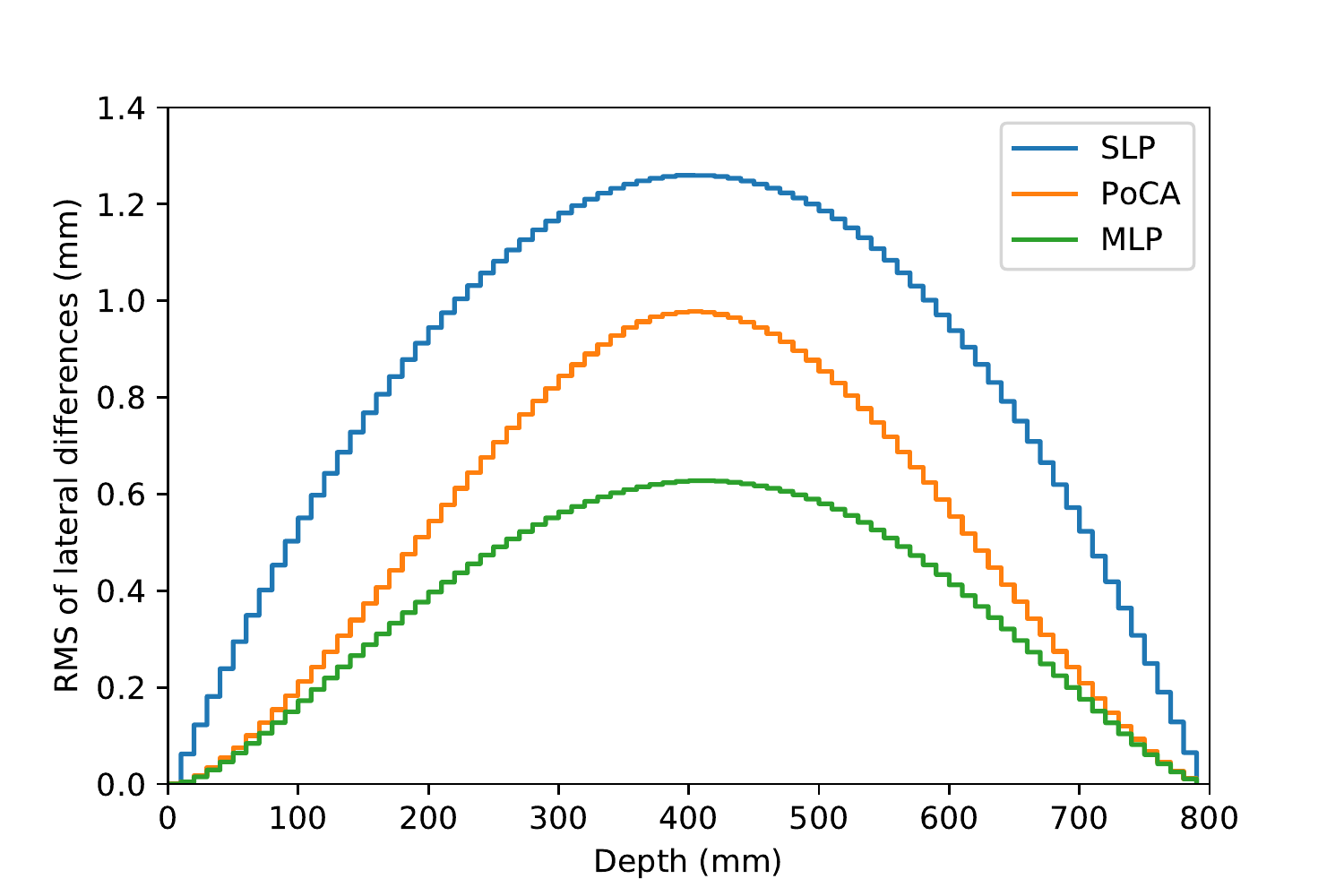}
  \caption{Precision of trajectory estimation algorithms}
  \label{fig:trackprecision}
\end{figure}

It is inappropriate to rank the precision of trajectory estimation algorithms by several examples like Figure~\ref{fig:trackdrawing} due to the stochastic MCS processes. 
Differences of simulated and estimated trajectories were calculated and then root mean square (RMS) were summarized at each depth for 3 algorithms as shown in Figure~\ref{fig:trackprecision}. 
Since estimations start with the entry and exit data, deviations between true trajectories and estimated ones are small at two ends while 
reaching maximum in the middle of depth axis. 
It is suggested that among these algorithms MLP results in the most accurate estimations in regard to deviation from simulated trajectories. 

\subsection{Reconstructed Images}

Image reconstruction algorithms were tested on simulation dataset generated with the geometry shown in Figure~\ref{fig:geant4geo}. 
For comparison, PoCA trajectory estimation were used in all three reconstruction algorithms, including direct allocation to PoCA (referred as ``PoCA'' in this section for simplicity), MLSD and MAP. 
In our tests, problem arose that reconstructed scattering densities from cosmic muons didn't meet values predicted from theory e.g. in Figure~\ref{fig:scatdensity}. 
The following contents will discuss the solution to this problem and comparison of reconstruction algorithms in terms of image quality. 

\subsubsection{Influence of Muon Momentum}

The idea of muon tomography relies on the differences in scattering density of materials. 
However, Rossi formula (Eq.~\ref{eqn:rossi}) suggests that muon momentum is also an important factor for determination of scattering densities. 
Muons from cosmic rays have kinetic energy in a wide range that slow muons tend to scatter at large angles. 

\begin{figure}[!htb]
  \centering
  \begin{subfigure}[t]{0.5\textwidth}
    \centering
    \includegraphics[width=0.9\hsize]{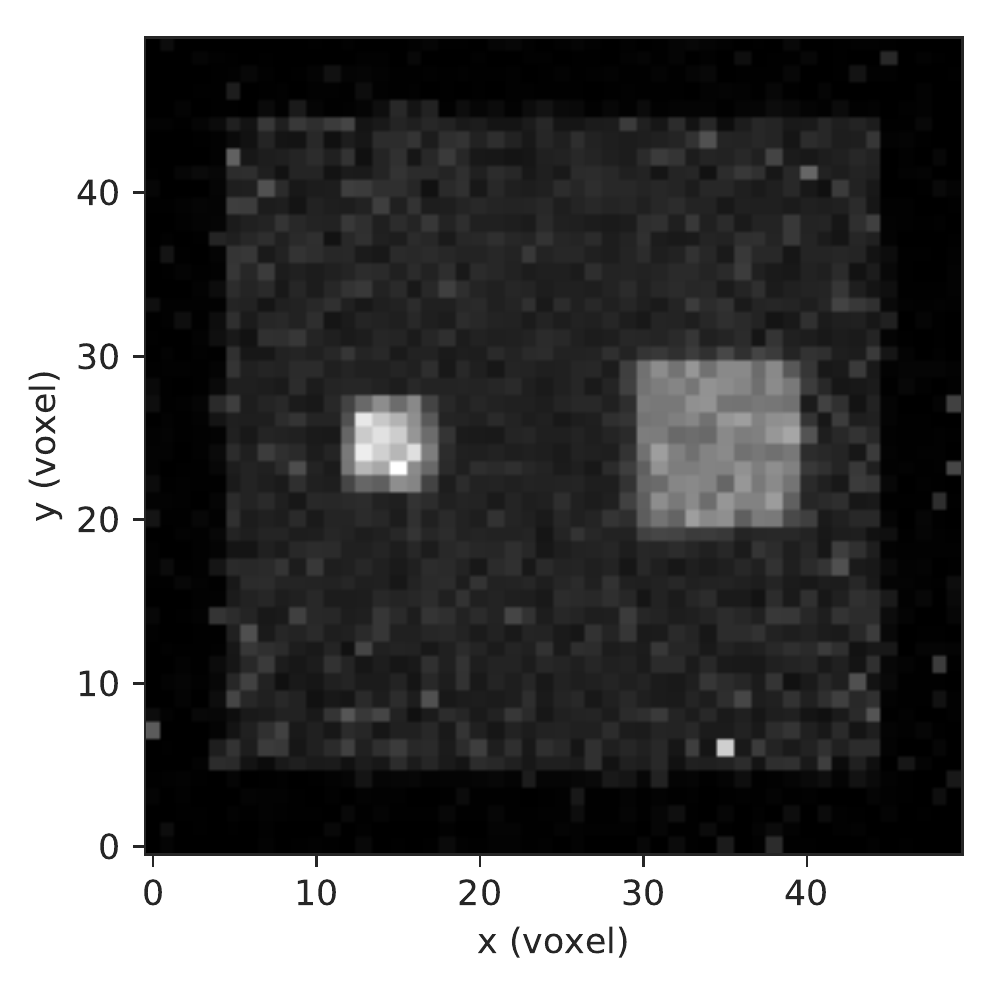}
    \caption{$E_k\leq 2 \mathrm{GeV}$}
  \end{subfigure}%
  \begin{subfigure}[t]{0.5\textwidth}
      \centering
      \includegraphics[width=0.9\hsize]{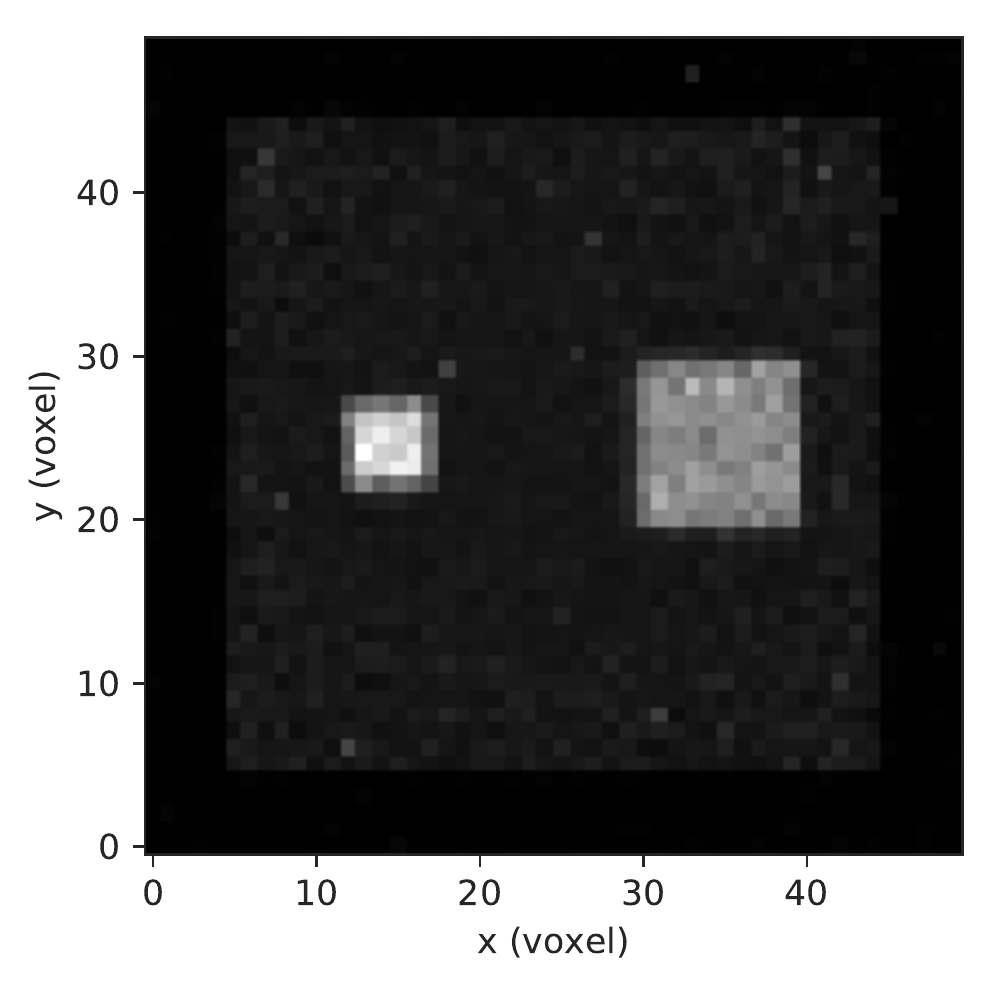}
      \caption{$E_k> 2 \mathrm{GeV}$}
  \end{subfigure}
  
  \caption{Horizontal projection of PoCA results from two groups of muons}
  \label{fig:pocalow}
\end{figure}

Since the dataset was generated from MC simulation, we have full control of incident muons and their properties. 
Dataset was divided into two groups by median muon kinetic energy of \SI{2}{GeV} and separately processes of image reconstruction were 
accomplished using PoCA algorithm. 
The results (Figure~\ref{fig:pocalow}) indicate that 
image reconstructed from high energy group is more clear and smooth while 
variation in scattering densities is a serious problem for muons of low kinetic energy.  

Exact momentum of incident muons are normally unknown in real tomography systems that 
representative values of scattering density should be evaluated by certain strategy. 

Define standard scattering density for muon of momentum $p_0$ as $\lambda_0$, then each scattering angle can be 
scaled to its equivalent with respect to standard muon momentum using $ \theta = \dfrac{p_0v_0}{pv} \theta_0 $. 
Then, the relation between measured scattering density and the standard $\lambda_0$ is derived in Eq.~\ref{eqn:pocaenergy}. 
The expectation value of ${\mathbb E}[\dfrac{1}{(pv)^2}]$ in the equation can be estimated since 
momentum distribution of cosmic muons is much easier to obtain than momentum of individual muon. 

\begin{equation}
  \label{eqn:pocaenergy}
  \begin{split}
    \sigma_\theta^2 =~& x \lambda = {\mathbb E}[\theta^2] \\
                    =~& {\mathbb E}[(\dfrac{p_0v_0}{pv})^2 \cdot \theta_0^2] = (p_0 v_0)^2 {\mathbb E}[\dfrac{1}{(pv)^2}] \sigma_0^2 \\ 
                    =~& (p_0 v_0)^2 {\mathbb E}[\dfrac{1}{(pv)^2}]\cdot  x\lambda_0  
  \end{split}
\end{equation}

\subsubsection{Comparing Reconstructed Images}

\begin{figure}[!htb]
  \centering
  \includegraphics
    [width=0.95\hsize]
    {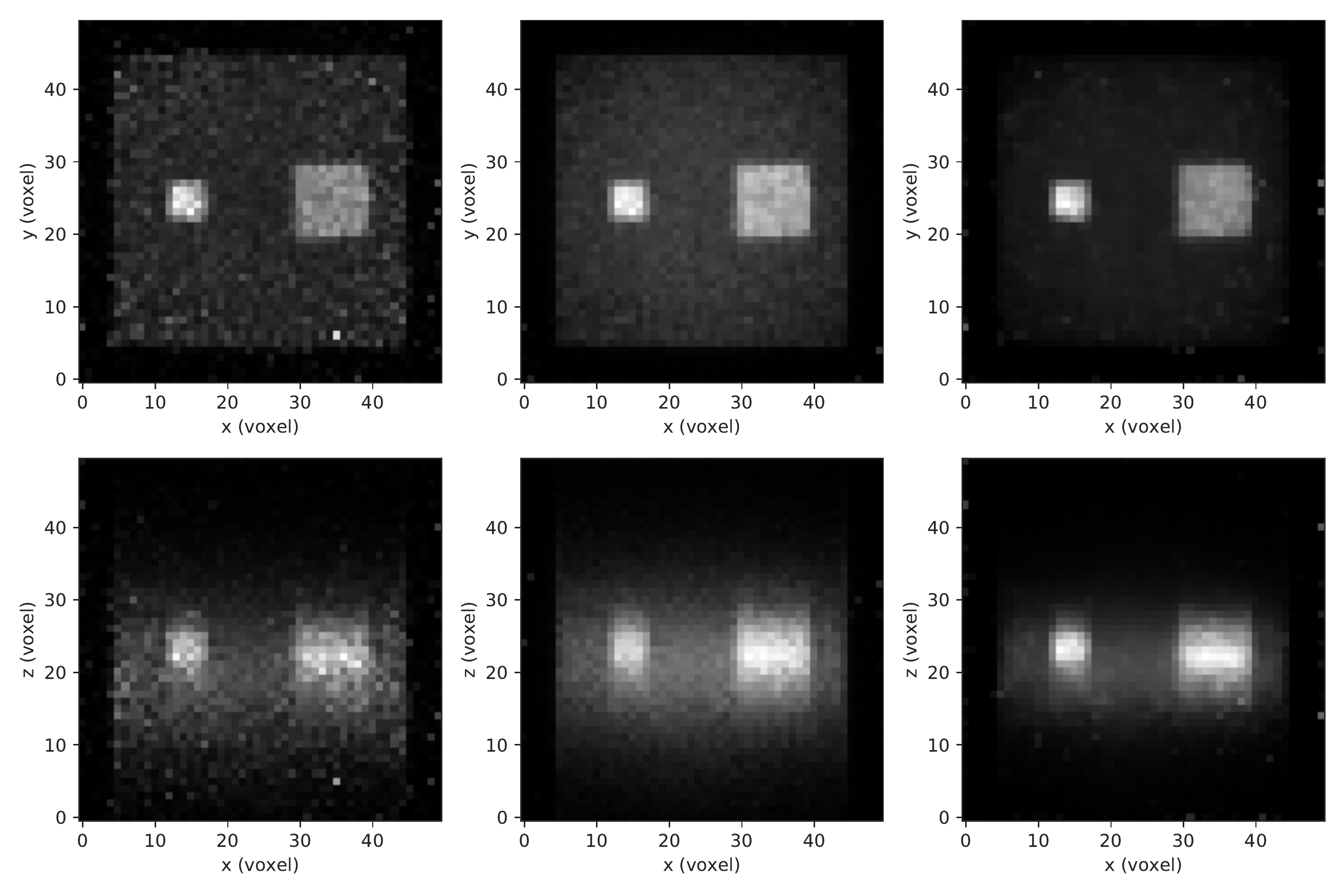}
  \caption{Comparison of Reconstruction Results. From left to right: PoCA algorithm, MLSD after 50 iterations, MAP with $l_1$ norm after 50 iterations. From top to bottom: horizontal plane and vertical plane views}
  \label{fig:compimg}
\end{figure}

Reconstructed images of PoCA, MLSD and MAP algorithms are projected to both horizontal and vertical planes and displayed in Figure~\ref{fig:compimg}. 
At first glance, the results indicates that muon tomography has much better resolution in horizontal plane than vertical axis. 
While all algorithms are sufficient to discover the scattering density differences among metal cubes and water container, 
image reconstructed by PoCA algorithm shows more noise than the other two statistical methods. 
Image from MAP algorithm preserves sharper edges than MLSD after the same number of iterations which is a reasonable consequence from regularization. 

\begin{figure}[!htb]
  \centering
  \includegraphics
    [width=\hsize]
    {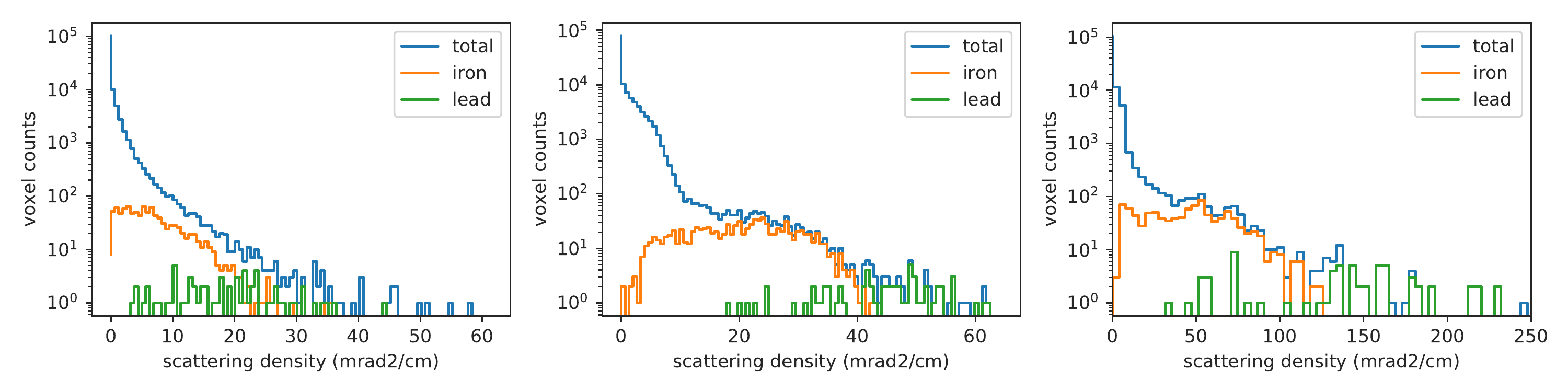}
  \caption{Comparison of scattering densities in different materials, from left to right: PoCA algorithm, MLSD after 50 iterations, MAP with $l_1$ norm after 50 iterations}
  \label{fig:compval}
\end{figure}

Image qualities are not always related to material discrimination abilities, thus, reconstructed scattering density values are 
statistical analysed within regions of two metal cubes and shown as histograms in Figure~\ref{fig:compval}. 
In PoCA algorithm, there are large scattering density values in voxels that do not belong to any metal cubes, 
which directly introduces the noise observed in reconstructed image. 
Part of voxels in iron cube render as smaller scattering density values in MAP algorithm with $l_1$ norm. 
Lead cube in all situations has voxels that infiltrate into the region belonging to iron materials. 
Overall, MLSD exhibits the best performance in terms of material discrimination though not of the best image quality compared to MAP. 

\section{Conclusion}\label{sec:discuss}

Theory and algorithms of image reconstruction in muon tomography were reviewed in this paper. 
We summarized the procedure of image reconstruction into two stages, 
upon which most of the state-of-the-art algorithms have been developed, as listed in Table~\ref{tab:summary_alg}. 

For the representative of simulated muon scattering data, physics models of MCS in GEANT4 toolkit were tested with theoretical formulas (Figure~\ref{fig:geant4model}). 
It is suggested that Wentzel model agrees with MCS theory better and should be used in simulation. 

Trajectory estimation algorithms were described in detail and their precision were evaluated with respect to simulation trajectories. 
It turns out that MLP is the most accurate trajectory estimation algorithm (Figure~\ref{fig:trackprecision}), 
however, the application of MLP in reconstruction remains an topic to be explored. 

Popular algorithms of muon tomography were implemented and tested on our simulation dataset, including PoCA, MLSD and MAP algorithms. 
Correction of scattering density considering the muon momentum should be applied in PoCA algorithm for comparable values with theoretical predictions. 
With more complicated algorithms that would bias the scattering density values in the future, some sort of calibration process for materials of interest should be applied experimentally. 
It is also implied in our results of dividing muons by 2 GeV threshold (Figure~\ref{fig:pocalow})
that momentum measurement of incident muon would greatly improve the reconstructed images. 

Though simple in design and programming, PoCA algorithm exhibits stable performance and is widely used as benchmark algorithm in experiments presented in literatures. 
Statistical methods like MLSD and MAP are superior in terms of image qualities and material discrimination abilities with the sacrifice of computing time and complexity in choices of parameters, such as regularization functions and iteration steps. 
There are promising new ideas in active research field for X-ray CT that would benefit muon tomography as well. 
For instance, regularization strategies in low-dose x-ray CT should be inspiring for muon tomography suffering from low counting statistics. 

Despite the non-stopping improvements of image reconstruction algorithms, whether muon tomography is a practical and compatible for  
industrial applications like cargo inspection or nuclear material discrimination depends heavily on the physics detection limits. 
Thus, a theoretical framework of muon tomography is also important and should be constructed to guide the future development of algorithms. 

\bibliographystyle{JHEP}
\bibliography{muon_tomography_review}

\providecommand{\href}[2]{#2}\begingroup\raggedright\begin{thebibliography}{10}

\bibitem{Patrignani2016}
{C. Patrignani, et al. (Particle Data Group)}, \emph{Review of particle
  physics},  vol.~40, p.~100001.
\newblock {IOP} Publishing, oct, 2016.
\newblock \href{http://dx.doi.org/10.1088/1674-1137/40/10/100001}{DOI}.

\bibitem{gaisser_engel_resconi_2016}
T.~K. Gaisser, R.~Engel and E.~Resconi, \emph{Particle physics}, p.~30–64.
\newblock Cambridge University Press, 2~ed., 2016.
\newblock 10.1017/CBO9781139192194.005.

\bibitem{Shukla2018}
P.~Shukla and S.~Sankrith, \emph{{Energy and angular distributions of
  atmospheric muons at the Earth}},
  \href{http://dx.doi.org/10.1142/S0217751X18501750}{\emph{International
  Journal of Modern Physics A} {\bfseries 33} (2018) },
  [\href{https://arxiv.org/abs/arXiv:1606.06907v3}{{\ttfamily
  arXiv:1606.06907v3}}].

\bibitem{Rossi1941}
B.~Rossi and K.~Greisen, \emph{{Cosmic-ray theory}},
  \href{http://dx.doi.org/10.1103/RevModPhys.13.240}{\emph{Reviews of Modern
  Physics} {\bfseries 13} (1941) 240--309}.

\bibitem{Moliere1948}
G.~Moliere, \emph{{Theorie der Streuung schneller geladener Teilchen II
  Mehrfach-und Vielfachstreuung}},
  \href{http://dx.doi.org/10.1515/zna-1948-0203}{\emph{Zeitschrift fur
  Naturforschung - Section A Journal of Physical Sciences} {\bfseries 3} (1948)
  78--97}.

\bibitem{Highland1975}
V.~L. Highland, \emph{{Some practical remarks on multiple scattering}},
  \href{http://dx.doi.org/10.1016/0029-554X(75)90743-0}{\emph{Nuclear
  Instruments and Methods} {\bfseries 129} (1975) 497--499}.

\bibitem{Lynch1991}
G.~R. Lynch and O.~I. Dahl, \emph{{Approximations to multiple Coulomb
  scattering}},
  \href{http://dx.doi.org/10.1016/0168-583X(91)95671-Y}{\emph{Nuclear Inst. and
  Methods in Physics Research, B} {\bfseries 58} (1991) 6--10}.

\bibitem{Procureur2018}
S.~Procureur, \emph{{Muon imaging : Principles , technologies and
  applications}},
  \href{http://dx.doi.org/10.1016/j.nima.2017.08.004}{\emph{Nuclear Inst. and
  Methods in Physics Research, A} {\bfseries 878} (2018) 169--179}.

\bibitem{Borozdin2003}
K.~N. Borozdin, G.~E. Hogan, C.~Morris, W.~C. Priedhorsky, A.~Saunders, L.~J.
  Schultz et~al., \emph{{Surveillance: Radiographic imaging with cosmic-ray
  muons.}}, \href{http://dx.doi.org/10.1038/422277a}{\emph{Nature} {\bfseries
  422} (2003) 277--278}.

\bibitem{Schultz2003}
L.~Schultz, \emph{{COSMIC RAY MUON RADIOGRAPHY}}.
\newblock PhD thesis, Portland State University, 2003.

\bibitem{Pesente2009}
S.~Pesente, S.~Vanini, M.~Benettoni, G.~Bonomi, P.~Calvini, P.~Checchia et~al.,
  \emph{{First results on material identification and imaging with a
  large-volume muon tomography prototype}},
  \href{http://dx.doi.org/10.1016/j.nima.2009.03.017}{\emph{Nuclear Instruments
  and Methods in Physics Research, Section A: Accelerators, Spectrometers,
  Detectors and Associated Equipment} {\bfseries 604} (2009) 738--746}.

\bibitem{Wang2012}
X.~Wang, J.~Cheng, Y.~Wang, Q.~Yue, Z.~Zhao, Z.~Zeng et~al., \emph{{Design and
  construction of muon tomography facility based on MRPC detector for high-Z
  materials detection}},
  \href{http://dx.doi.org/10.1080/11356405.2015.1120451}{\emph{IEEE Nuclear
  Science Symposium Conference Record} (2012) 83--85}.

\bibitem{Jonkmans2013}
G.~Jonkmans, V.~N. Anghel, C.~Jewett and M.~Thompson, \emph{{Nuclear waste
  imaging and spent fuel verification by muon tomography}},
  \href{http://dx.doi.org/10.1016/j.anucene.2012.09.011}{\emph{Annals of
  Nuclear Energy} {\bfseries 53} (2013) 267--273}.

\bibitem{Clarkson2014}
A.~Clarkson, D.~J. Hamilton, M.~Hoek, D.~G. Ireland, J.~R. Johnstone, R.~Kaiser
  et~al., \emph{{The design and performance of a scintillating-fibre tracker
  for the cosmic-ray muon tomography of legacy nuclear waste containers}},
  \href{http://dx.doi.org/10.1016/j.nima.2014.02.019}{\emph{Nuclear Instruments
  and Methods in Physics Research, Section A: Accelerators, Spectrometers,
  Detectors and Associated Equipment} {\bfseries 746} (2014) 64--73},
  [\href{https://arxiv.org/abs/1309.3400}{{\ttfamily 1309.3400}}].

\bibitem{Anghel2015}
V.~Anghel, J.~Armitage, F.~Baig, K.~Boniface, K.~Boudjemline, J.~Bueno et~al.,
  \emph{{A plastic scintillator-based muon tomography system with an integrated
  muon spectrometer}},
  \href{http://dx.doi.org/10.1016/j.nima.2015.06.054}{\emph{Nuclear Instruments
  and Methods in Physics Research, Section A: Accelerators, Spectrometers,
  Detectors and Associated Equipment} {\bfseries 798} (2015) 12--23}.

\bibitem{Luo2016}
Z.~Luo, X.~Wang, Z.~Zeng, Y.~Wang, M.~Zeng, J.~Cheng et~al., \emph{{Energy
  measurement and application on material discrimination in muon tomography}},
  \href{http://dx.doi.org/10.1109/NSSMIC.2015.7581957}{\emph{2015 IEEE Nuclear
  Science Symposium and Medical Imaging Conference, NSS/MIC 2015} (2016) 2--5}.

\bibitem{Schulte2008}
R.~W. Schulte, S.~N. Penfold, J.~T. Tafas and K.~E. Schubert, \emph{{A maximum
  likelihood proton path formalism for application in proton computed
  tomography}}, \href{http://dx.doi.org/10.1118/1.2986139}{\emph{Medical
  Physics} {\bfseries 35} (2008) 4849--4856}.

\bibitem{Yi2016}
H.~Yi, Z.~Zeng, B.~Yu, J.~Cheng, Z.~Zhao, X.~Wang et~al.,
  \emph{{Bayesian-theory-based most probable trajectory reconstruction
  algorithm in cosmic ray muon tomography}},
  \href{http://dx.doi.org/10.1109/NSSMIC.2014.7431084}{\emph{2014 IEEE Nuclear
  Science Symposium and Medical Imaging Conference, NSS/MIC 2014} (2016) 1--4}.

\bibitem{Chatzidakis2018}
S.~Chatzidakis, Z.~Liu, J.~P. Hayward and J.~M. Scaglione, \emph{{A generalized
  muon trajectory estimation algorithm with energy loss for application to muon
  tomography}}, \href{http://dx.doi.org/10.1063/1.5024671}{\emph{Journal of
  Applied Physics} {\bfseries 123} (2018) }.

\bibitem{Schultz2007a}
L.~J. Schultz, G.~S. Blanpied, K.~N. Borozdin, A.~M. Fraser, N.~W. Hengartner,
  A.~V. Klimenko et~al., \emph{{Statistical reconstruction for cosmic ray muon
  tomography}}, \href{http://dx.doi.org/10.1109/TIP.2007.901239}{\emph{IEEE
  Transactions on Image Processing} {\bfseries 16} (2007) 1985--1993}.

\bibitem{Liu2018}
Z.~Liu, S.~Chatzidakis, J.~M. Scaglione, C.~Liao, H.~Yang and J.~P. Hayward,
  \emph{{Muon Tracing and Image Reconstruction Algorithms for Cosmic Ray Muon
  Computed Tomography}},
  \href{http://dx.doi.org/10.1109/TIP.2018.2869667}{\emph{IEEE Transactions on
  Image Processing} {\bfseries 28} (2018) 426--435}.

\bibitem{Yu2016}
B.~Yu, Z.~Zhao, X.~Wang, D.~Wu, Z.~Zeng, M.~Zeng et~al., \emph{{A unified
  framework for penalized statistical muon tomography reconstruction with edge
  preservation priors of lp norm type}},
  \href{http://dx.doi.org/10.1016/j.nima.2015.09.113}{\emph{Nuclear Instruments
  and Methods in Physics Research, Section A: Accelerators, Spectrometers,
  Detectors and Associated Equipment} {\bfseries 806} (2016) 199--205}.

\bibitem{Siddon1985}
R.~L. Siddon, \emph{{Fast calculation of the exact radiological path for a
  3-dimensional CT array}}, {\emph{Medical Physics Med. Phys. Med. Phys}
  {\bfseries 12} (1985) }.

\bibitem{Jacobs1998}
F.~Jacobs, \emph{{A fast algorithm to calculate the exact radiological path
  through a pixel or voxel space}}, .

\bibitem{Zhang2018}
H.~Zhang, J.~Wang, D.~Zeng, X.~Tao and J.~Ma, \emph{{Regularization strategies
  in statistical image reconstruction of low-dose x-ray CT: A review}},
  \href{http://dx.doi.org/10.1002/mp.13123}{\emph{Medical Physics} {\bfseries
  45} (2018) e886--e907}.

\bibitem{AGOSTINELLI2003250}
S.~Agostinelli, J.~Allison, K.~Amako, J.~Apostolakis, H.~Araujo, P.~Arce
  et~al., \emph{Geant4—a simulation toolkit},
  \href{http://dx.doi.org/https://doi.org/10.1016/S0168-9002(03)01368-8}{\emph{Nuclear
  Instruments and Methods in Physics Research Section A: Accelerators,
  Spectrometers, Detectors and Associated Equipment} {\bfseries 506} (2003) 250
  -- 303}.

\bibitem{CRY2007}
C.~{Hagmann}, D.~{Lange} and D.~{Wright}, \emph{Cosmic-ray shower generator
  (cry) for monte carlo transport codes},  in \emph{2007 IEEE Nuclear Science
  Symposium Conference Record}, vol.~2, pp.~1143--1146, Oct, 2007.
\newblock \href{http://dx.doi.org/10.1109/NSSMIC.2007.4437209}{DOI}.

\end{thebibliography}\endgroup

\end{document}